\documentclass[%
 aip,
 amsmath,amssymb,
reprint,%
]{revtex4-2}
\usepackage{graphicx}
\usepackage{dcolumn}
\usepackage{bm}
\usepackage{color}

\usepackage[utf8]{inputenc}
\usepackage[T1]{fontenc}
\usepackage{mathptmx}
\usepackage{subfig}
\usepackage{amsmath}
\usepackage[inline]{enumitem}
\usepackage{caption}
\usepackage{mathrsfs}
\usepackage{comment}

\usepackage{float}
\usepackage[normalem]{ulem}

\begin{document}

\title{Excitation of whistler and slow-X waves by runaway electrons in a collisional plasma}

\author{Qile Zhang}
\affiliation{Theoretical Division, Los Alamos National Laboratory, Los Alamos, NM 87545, USA}
\affiliation{University of Maryland, College Park, Maryland, 20742, USA}

\author{Yanzeng Zhang}
\affiliation{Theoretical Division, Los Alamos National Laboratory, Los Alamos, NM 87545, USA}
\affiliation{School of Nuclear Science and Technology, University of Science and Technology of China, Hefei, Anhui 230027, China}

\author{Xian-Zhu Tang}
\affiliation{Theoretical Division, Los Alamos National Laboratory, Los Alamos, NM 87545, USA}

\begin{abstract}
  Runaway electrons are known to provide robust ideal or collisionless
  kinetic drive for plasma wave instabilities in both the whistler and
  slow-X branches, via the anomalous Doppler-shifted cyclotron
  resonances.  In a cold and dense post-thermal-quench plasma,
  collisional damping of the plasma waves can compete with the
  collisionless drive. Previous studies have found that for its higher
  wavelength and frequency, slow-X waves suffer stronger collisional
  damping than the whistlers, while the ideal growth rate of slow-X
  modes is higher. Here we study runaway avalanche distributions that maintain the same eigen distribution and increase only in magnitude over time. The distributions are computed from the
  relativistic Fokker-Planck-Boltzmann solver, upon which a linear
  dispersion analysis is performed to search for the most unstable or
  least damped slow-X and whistler modes. Taking into account the
  effect of plasma density, plasma temperature, and effective charge
  number, we find that the slow-X modes tend to be excited before the
  whistlers in a runaway current ramp-up. Furthermore, even when the
  runaway current density is sufficiently high that both branches are
  excited, the most unstable slow-X mode has much higher growth rate
  than the most unstable whistler mode. The qualitative and
  quantitative trends uncovered in current study indicate that even
  though past experiments and modeling efforts have concentrated on whistler modes, there's a compelling case that slow-X modes should also be a key area of focus.
\end{abstract}

\maketitle

\section{Introduction}

During both tokamak
startup~\cite{deVries-nf-2019,deVries-ppcf-2020,Hoppe-etal-JPP-2022}
and major
disruptions,~\cite{Hender:2007,Boozer:2015,breizman2019physics}
runaway acceleration of electrons along the magnetic field can produce
a relativistic tail distribution that carries most of the plasma
current.  Depending on how strong the parallel electric field
$E$ is compared with the Dreicer field
($E_D$),~\cite{Dreicer59} the avalanche threshold electric field
($E_{AV}$),~\cite{Rosenbluth97,Aleynikov-Breizman-PRL15,mcdevitt-etal-ppcf-2018}
and the Connor-Hastie threshold electric field
($E_C$),~\cite{connor-hastie-NF75} the runaway electrons can obtain a
slide-away distribution, an avalanche distribution, and a primary
distribution.  Specifically, for $E \sim E_D,$ one has a
slide-away distribution. For $E_{AV} < E \ll E_D,$ an
avalanche distribution is produced as a result of the knock-on
collisions between primary runaways and the background
electrons.~\cite{Sokolov:1979,Jayakumar:1993,Rosenbluth97} For $E_c <
E < E_{AV},$ there is no exponential growth of the secondary
runaways and the runaway distribution is primarily made of primary
runaways. Mathematically, both the primary and slide-away
distributions are solved from the relativistic Fokker-Planck
equation,~\cite{guo2017phase} while the avalanche distribution is the
solution of the more complicated relativistic Fokker-Planck-Boltzmann
equation,~\cite{,McDevitt-2019,guo-mcdevitt-tang-pop-2019} where the
Boltzmann collision operator is required for accurately describing the
physics of large-angle collisions.

A high energy tail of electrons, with narrow pitch spread with respect
to the magnetic field, resembles a high-energy beam in a background
plasma, which is known to excite plasma
waves~\cite{Parail-NF-1978,pokol-etal-PoP-2014,Aleynikov2015NucFu}. These
plasma waves can in turn modify the runaway distribution, for example,
through effective pitch angle scattering via collisionless
wave-particle interactions studied using the quasilinear diffusion
formalism~\cite{guo2018control,Liu2018prl,Pokol2014PhPl} or nonlinear
kinetic simulations.~\cite{zhang-etal-arXiv-2024} If the background
plasma is relatively cold, which is certainly the case in general for
a post-thermal-quench tokamak plasma, collisional damping can be substantial
so it can significantly lower the growth rate of an ideal instability
and may be able to completely stabilize the modes.~\cite{Aleynikov2015NucFu}

The ideal or collisionless drive for the wave instability comes from
the tail electron distribution by the way of a Doppler-shifted
cyclotron resonance,
\begin{align}
\omega - k_\parallel v \xi - n \omega_{ce}/\gamma = 0,\label{eq:resonance-condition}
\end{align}
where the wave has frequency $\omega$ and parallel wave vector
$k_\parallel,$ the runaway electron has a Lorentz factor
$\gamma=1/\sqrt{1-v^2/c^2}$ and speed $v$ with pitch $\xi \equiv
cos\theta= v_\parallel/v,$ the signed non-relativistic electron
cyclotron frequency is $\omega_{ce} = eB/mc < 0$ for the negative
electron charge $e.$ Here $n$ is a non-negative integer, with $n=0$
denoting the Cherenkov resonance and $n=1,2,\cdots$ for various
anomalous Doppler-shifted cyclotron resonances. The intersection point of the
resonance curve, Eq.~(\ref{eq:resonance-condition}), with the plasma
wave dispersion $\omega(\mathbf{k})$ signify the corresponding plasma
waves of $(\omega_r, k_\parallel)$ that can be resonantly driven by
the runaways of energy $\gamma$ and pitch $\xi.$ The resonant wave
frequency $\omega_r$ has an additional dependence on $k_\perp$ through
the wave dispersion relation.  For a low-temperature magnetized plasma
where the wave dispersion is approximated by the cold plasma
dispersion relation, the resonance line can intercept both the
lower-frequency magnetosonic-whistler wave
branch~\cite{Fülöp2006pop,Liu2018prl}, and the extraordinary electron
wave branch,~\cite{Pokol2014PhPl, Komar2012JPhCS} also known as the
slow-X mode.\cite{Ram2000pop} An interesting finding from these
analyses~\cite{Pokol2014PhPl, Komar2012JPhCS} is that the slow-X
modes, of much higher frequencies than those of the whistler modes,
tend to have a much higher ideal growth rate and stronger quasilinear
pitch angle diffusion, compared with the whistler modes.
The important physics question of how these runaway-driven wave
instabilities are collisionally damped in a relatively cold plasma
found in disruptions and startups, requires a kinetic description that
was given by Aleynikov and Breizman in
Ref.~\onlinecite{Aleynikov2015NucFu}.  One key takeaway from the
damping rate calculation~\cite{Aleynikov2015NucFu} is that the
collisional damping rate can be substantially lower than the electron
collision rate. This
is particularly the case for low-frequency whistler branch, as labeled
by the solid segment of the low-frequency branch in the Fig.~2 of
Ref.~\onlinecite{Aleynikov2015NucFu}. In some previous literature,  the collisional damping rate was mistaken to
be the electron collision rate.~\cite{Fülöp2006pop,fulop-etal-pop-2009,brambilla-pop-1995}   From the same Fig.~2 of
Ref.~\onlinecite{Aleynikov2015NucFu}, another
impactful conclusion was drawn that in contrast to whistlers, the
collisional damping rate of the higher-frequency slow-X modes is a lot
higher, for example, by more than one order of magnitude or even
greater. The practical implication -- although not explicitly stated
in Ref.~\onlinecite{Aleynikov2015NucFu} but appearing to have received
wide acceptance as a corollary of the calculation in
Ref.~\onlinecite{Aleynikov2015NucFu} -- is that since the
high-frequency slow-X modes are strongly damped by collisions compared
with low-frequency whistlers, as the runaway current density builds up
after the plasma cools, it would be the low-frequency whistlers that
are preferentially excited. If this is true, the slow-X modes, despite
having much higher ideal growth rates, would only play a secondary
role, if any, in realistic experiments.

Here we revisit this issue by introducing a more realistic runaway
distribution in energy and pitch, in contrast to the previous
calculation that focused on an exponential model electron spectrum of
an averaged $\gamma\sim20$ and a constant pitch
spread.~\cite{Aleynikov2015NucFu} For definitiveness of the
calculation, we have focused on an avalanche distribution, which as
Rosenbluth and Putvinskii~\cite{Rosenbluth97} have previously shown,
has a characteristic distribution in energy and pitch
($\hat{f}_{RE}(p,\xi)$) for given background plasma parameters (e.g.,
density $n_e,$ temperature $T_e,$ and effective charge number
$Z_{eff}$) and parallel electric field ($E$). The runaway
distributions of different runaway current density during the
avalanche growth period can simply scale up from an eigenfunction
$\hat{f}_{RE}(p,\xi)$ by a factor $C,$
\begin{align}
f_{RE} = C \hat{f}_{RE}(p,\xi).
\end{align}
In this work, we will examine three prototypical avalanche
distributions: (1) $E = 65E_c$ and $Z_{eff}=1;$ (2) $E = 200 E_c$ and
$Z_{eff}=1;$ and (3) $E=200 E_c$ and $Z_{eff}=5.$ Case (2) has much
higher parallel electric field as a multiple of $E_c,$ so it is
expected to produce higher maximum momentum and smaller pitch spread
than case (1). For the larger $Z_{eff},$ case (3) is expected to have
a larger pitch spread than case (2).

Note that the ideal instability drive scales
linearly with the runaway current density, and the collisional damping
rate scales inversely with background electron temperature $T_e^{-3/2}.$ Consistent with these expectations, we
find that to excite either whistlers or slow-X  for lower $T_e$ (which
means stronger collisional damping), the runaway current density must
surpass a higher threshold, which we write $j_c^{WS}$ for whistlers
and $j_c^{SX}$ for slow-X modes.  However, our findings on the relative
magnitude of $j_c^{WS}$ and $j_c^{SX}$ contradict earlier
expectation. Specifically, the most unstable slow-X modes are
triggered first as the runaway current density ramps up in an
avalanche. In other words, the threshold runaway current density for
exciting whistlers, $j_c^{WS},$ is actually greater than that for
exciting slow-X modes, $j_c^{SX}.$ The gap between $j_c^{WS}$ and
$j_c^{SX}$ scales with the collisional damping rate, so it becomes larger
for lower $T_e.$ Note that we still get this same conclusion even with the exponential model distribution in Ref.~\onlinecite{Aleynikov2015NucFu}. In contrast to the focus in the literature on whistlers
in current experiments~\cite{Spong2018prl,Liu2018prl} and in
anticipation of future reactor-scale
experiments,~\cite{Aleynikov2015NucFu} our calculation indicates that
the slow-X mode is likely the first to be excited. In addition,
because its growth rate far outpaces that of whistlers as the runaway
current density ramps up further in an avalanche, the nonlinear saturation of
slow-X modes should dominate the self-mediation of runaways by
self-excited waves. This actually brings additional difficulties in
experimental diagnosis since we have found that the nonlinear
saturation of slow-X modes involves whistler waves via parametric decay
and secondary/tertiary instabilities,~\cite{zhang-etal-arXiv-2024} so
one must be prepared to experimentally separate the roles of
runaway-driven primary whistlers and the slow-X induced
secondary/tertiary whistlers.

To ensure the accuracy of our calculation, we have also performed
benchmark studies with those of Ref.~\onlinecite{Aleynikov2015NucFu}
for the exponential model distribution, the details of which are shown
in the Appendix.  As a quick summary, the benchmark on the collisional
damping rates on the whistler and slow-X waves in resonance with a
runaway electron beam of $\gamma=20$ and $\xi=1$ is in excellent
agreement with Ref.~\onlinecite{Aleynikov2015NucFu}. But the ideal
growth rate benchmark for a model runaway distribution with finite
energy and pitch spread ($\gamma_0=25$ and $\theta_0=0.1$) shows
appreciable difference for both whistlers and the magnetized plasma
wave.  To shed additional light on this discrepancy, we proceed to
perform a higher-order expansion to obtain an analytical result of the
whistler growth rate. Our numerically calculated growth rates of the
whistler modes were found to be in excellent agreement with the
analytical results.  Since Ref.~\onlinecite{Aleynikov2015NucFu} did
not evaluate the ideal growth rate of slow-X modes, a direct benchmark
on slow-X modes, which would be informative, is not possible. We
instead benchmark with another publication Ref.~\onlinecite{Komar2012JPhCS} in
the Appendix.

The rest of the paper is organized as
follows. Section~\ref{sec:formulation} gives the theoretical
formulation for the linear dispersion analysis of the runaway-driven
whistler and slow-X modes. Section \ref{sec:mode-scan} uses this
analysis to scan for the most unstable modes for the calculated
avalanche distributions that depend on the electric field,
temperature, density and effective charge. Section
\ref{sec:conclusion} draws the conclusion.

\section{Formulation of the linear dispersion analysis\label{sec:formulation}}

\subsection{Collisional damping by the background plasma of whistler and slow-X waves}

The dispersion relation of plasma wave with frequency $\omega$ and wave vector $\{k_\alpha\}$ for
the cold magnetized plasma is solved from the linearized wave equation
\begin{align}
  \left[k_\alpha k_\beta c^2 - \delta_{\alpha\beta} k^2c^2 + \omega^2\epsilon_{\alpha\beta}\right] E_\beta =
  0 \label{eq:linearized-wave-eq}
\end{align}
by finding the roots $\omega(k_\alpha)$ of
\begin{align}
  \textrm{det}\left[ N_\alpha N_\beta - \delta_{\alpha\beta} N^2 + \epsilon_{\alpha\beta}\right] = 0.
  \label{eq:cold-plasma-dispersion}
\end{align}
Here
\begin{align}
  \epsilon_{\alpha\beta} (\omega) \equiv \epsilon\delta_{\alpha\beta} + ig e_{\alpha\beta\gamma} b_\gamma
  + \left(\eta - \epsilon\right) b_\alpha b_\beta \label{eq:dielectric-tensor}
\end{align}
is a dielectric tensor, $k_\alpha$ is the wave vector, $E_\beta$ the
polarization vector, $N_\alpha\equiv k_\alpha c/\omega$ the refractive
index, and $b_\alpha\equiv B_\alpha/B$ the component of a unit vector
along the magnetic field.
In the collisionless limit, the dielectric tensor is Hermitian with components
\begin{align}
  \epsilon & = \epsilon^H \equiv 1 - \sum_{p\in \{e,i\}}\frac{\omega_{p,s}^2}{\omega^2-\omega_{c,s}^2},
  \label{eq:epsilon}\\
  g & = g^H \equiv - \sum_{s\in \{e,i\}} \frac{\omega_{c,s}}{\omega} \frac{\omega_{p,s}^2}{\omega^2 - \omega_{c,s}^2},
  \label{eq:g} \\
  \eta & = \eta^H \equiv 1 - \sum_{s\in\{e,i\}} \frac{\omega_{p,s}^2}{\omega^2}. \label{eq:eta}
\end{align}
Here the species summation is over both the electrons and ions in the plasma. 

The polarization vector of the wave dispersion from
Eq.~(\ref{eq:linearized-wave-eq},\ref{eq:cold-plasma-dispersion}), in the case of a
uniform magnetic field in the $z$ coordinate direction,
has the form
\begin{align}
  \mathbf{E} & = \frac{E_x\left(\mathbf{N} - \mathbf{b}\left(\mathbf{N}\cdot\mathbf{b}\right)\right) + E_y\left[\mathbf{b}\times\mathbf{N}\right]}{\sqrt{\mathbf{N}^2 - \left(\mathbf{N}\cdot\mathbf{b}\right)^2}} + E_z \mathbf{b},\\
  E_x & = 1,\\
  E_y & = i\frac{g}{\epsilon-N^2}, \\
  E_z &= - \frac{\mathbf{N}\cdot\mathbf{b}}{\eta - N^2 + \left(\mathbf{N}\cdot\mathbf{b}\right)^2}
  \sqrt{N^2 - \left(\mathbf{N}\cdot\mathbf{b}\right)^2}.
\end{align}

\begin{figure}[H]
\centering
\includegraphics[width=0.5\textwidth]{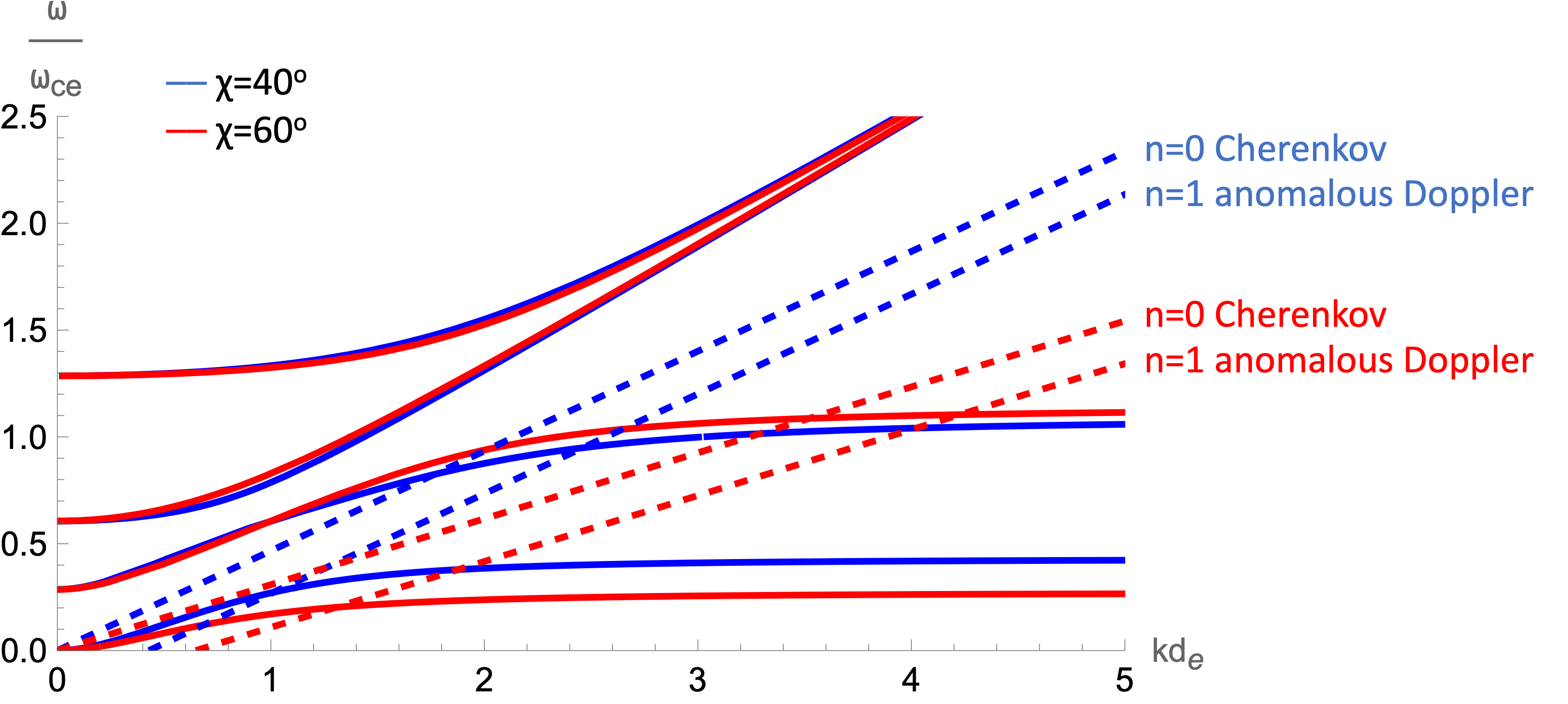}
\caption{Dispersion relations (solid) and sample Cheronkov and anomalous Doppler resonance conditions (dash, based on a parallel beam of $\gamma=5$) for two example propagation angles $\chi$ (red and blue), showing resonance occurring at very different $k$ and $\omega$ at different branches and $\chi$. $|\omega_{ce}|/\omega_{pe}=1.65$.  }
\label{fig:dispersion-resonance}
\end{figure}

In Fig.~\ref{fig:dispersion-resonance}, we plot the cold plasma
dispersion $\omega(k)$ and the resonance conditions for both
the $n=0$ Cherenkov resonance and the $n=1$ anomalous Doppler-shifted
cyclotron resonance.  Both the low-frequency branch of
whistlers/magnetized plasma waves, and the high-frequency branch of
slow-X/upper hybrid waves, can intercept the resonance line of a
fixed $\gamma$ and pitch $\xi.$
For both branches, we show two values of $\cos\chi\equiv k_\parallel/k,$ which can intercept
the corresponding resonance lines at different $(\omega,k).$
This motivates the later study in which we must vary $\chi$
in the search of the most unstable slow-X and whistler modes.

Following Ref.~\onlinecite{Aleynikov2015NucFu}, the collisional damping rate is obtained by substituting
$\omega \rightarrow \omega + i\nu_e$ with
\begin{align}
  \nu_e = \frac{4\sqrt{2\pi} \ln\Lambda e^4}{3m_e^{1/2} T_e^{3/2}} \sum_Z Z^2 n_Z \label{eq:nu-e}
\end{align}
the electron-ion collision frequency, into the
conductivity tensor only.
In the regime of interest to us, which has $\omega\gg \nu_e,$ the collisions add a small anti-Hermitian component
$\epsilon_{\alpha\beta}^A$ to the dielectric tensor, so
Eqs.~(\ref{eq:epsilon},\ref{eq:g},\ref{eq:eta}) now take the form
\begin{align}
\epsilon = \epsilon^H + \epsilon^A, g = g^H + g^A, \eta = \eta^H + \eta^A,
\end{align}
with the anti-Hermitian parts given by
\begin{align}
  \epsilon^A & = i\frac{\nu_e}{\omega} \frac{\omega_{pe}^2\omega^2 + \omega_{ce}^2\omega_{pe}^2}{\left(\omega^2-\omega_{ce}^2\right)^2}, \\
    g^A &= 2i\frac{\nu_e}{\omega} \frac{\omega_{pe}^2\omega\omega_{ce}}{\left(\omega^2-\omega_{ce}^2\right)^2}, \\
    \eta^A &= i\frac{\nu_e}{\omega}\frac{\omega_{pe}^2}{\omega^2}.
\end{align}
Treating this as a perturbation to the linearized wave equation, one finds
the collisional damping rate~\cite{Aleynikov2015NucFu},
\begin{align}
\Gamma_\nu = - i\frac{E_\alpha^*E_\beta \omega^2\epsilon_{\alpha\beta}^A}{E_\alpha^*E_\beta\frac{\partial}{\partial\omega} \omega^2 \epsilon_{\alpha\beta}^H} \label{eq:gamma-nu}
\end{align}
with $E_\alpha$ the wave polarization vector previously given.

\begin{widetext}
\begin{figure}[H]
\centering
\includegraphics[width=0.9\textwidth]{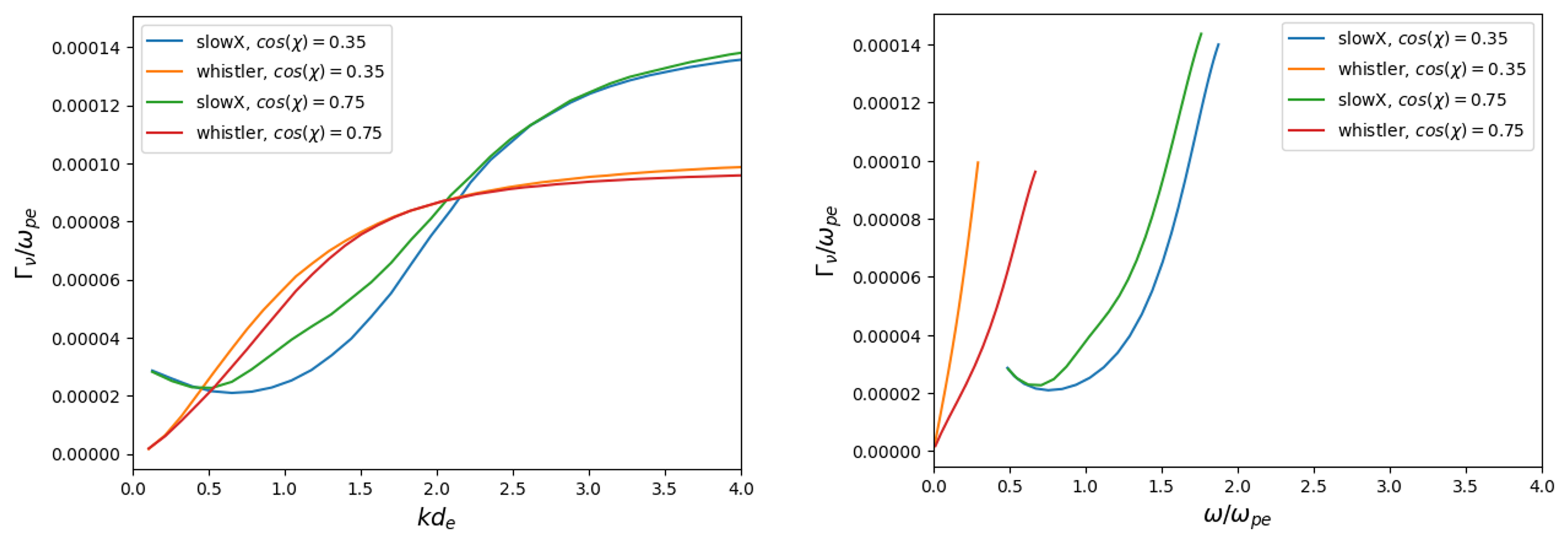}
\caption{Wave damping as a function of $k$ and $\omega$ to show the nonlinear scaling. The calculation is based on $|\omega_{ce}|/\omega_{pe}=1.65$, density $n_e=10^{20}m^{-3}$, temperature $T_e=10$~eV and $Z_{eff}=1$.  }
\label{fig:wave-damping}
\end{figure}
\end{widetext}

In Fig.~\ref{fig:wave-damping}, we plot the collisional wave damping
rates as a function of $k$ and $\omega$ on the whistlers and slow-X branches at two different example propagation angles. The damping rates increase nonlinearly as a function of $\omega$ and $k$, and the actual damping highly depends on the specific resonant wave mode.  One would need to scan the propagation
angle $\chi$ and $k$ (or $\omega$) to search for the most unstable waves under these
damping rates, using the avalanche runaway distributions that will be calculated from the
Fokker-Planck-Boltzmann solver in sec.~\ref{sec:mode-scan}.

\subsection{Growth rate of runaway-driven plasma wave instability in a relatively cold plasma}

For the runaway electron driven wave instabilities, we follow the
standard perturbative analysis that treats the runaway electron
component ($f_{RE}$) as a small perturbation to the background
distribution ($f_M\left(n_0,T_0\right)$),
\begin{align}
  f_e(p,\xi) = f_M(n_0,T_0) + f_{RE}(p,\xi).
\end{align}
By small, we mean the runaway electron density $n_b$ is much lower
than the background electron density $n_0,$ i.e., $n_b \ll n_0.$
Similarly we treat the runaway modification of the plasma dispersion
as a small perturbation, $\omega = \omega_0 + \delta \omega,$ where
$\omega_0(\mathbf{k})$ is the plasma dispersion in the absence of the
runaway component.
The ideal growth rate of the runaway excited wave instability would simply
be the imaginary part of $\delta\omega,$ i.e., $\Gamma_b = \textrm{Im}\left(\delta\omega\right).$
For our purpose, the background plasma is cold $f_M(n_0,T_0=0),$ and the $\omega_0(\mathbf{k})$
is the cold plasma dispersion we have already introduced at the beginning of this section.
The ideal growth rate is found from \cite{Aleynikov2015NucFu}
\begin{widetext}
\begin{align}
\Gamma_\text{b} =\;&
4\pi^2 e^2 \int \mathrm{d}p\, \mathrm{d}\theta\, (2\pi\, p^2 \sin\theta)
\sum_{n=-\infty}^{\infty} Q_n
\left[
  \left(
    V \frac{\partial F_\text{b}}{\partial p}
    + \frac{V}{p} \frac{n\omega_\text{cb} - \omega \sin^2\theta}
                         {\omega \cos\theta \sin\theta}
      \frac{\partial F_\text{b}}{\partial \theta}
  \right)
  \delta\left(\omega - k_\parallel V \cos\theta - n\omega_\text{cb}\right)
\right] \notag\\
&\times
\left[
  \left(1 - E_y^2\right)
  \frac{1}{\omega} \frac{\partial}{\partial\omega} \omega^2 \varepsilon
  + 2i E_y
  \frac{1}{\omega} \frac{\partial}{\partial\omega} \omega^2 g
  + E_z^2
  \frac{1}{\omega} \frac{\partial}{\partial\omega} \omega^2 \eta
\right]^{-1},
  \label{eq:ideal-growth-rate}
\end{align}
\end{widetext}
where
\begin{align}
Q_n \equiv \left\{
  \frac{n \omega_\text{cb}}{k_\perp V} J_n
  + E_z \cos\theta\, J_n
  + i E_y \sin\theta\, J_n'
\right\}^2.
\end{align}
The argument of the Bessel function $J_n$ and its derivative $J_n'$ is
$k_\perp \rho=k_\perp V \sin\theta / \omega_{\text{cb}}$, where $V$ is
the particle velocity, $\omega_{\text{cb}} = \omega_{ce} / \gamma$ is
the gyro-frequency and $\rho=V \sin\theta / \omega_{\text{cb}}$ is the larmor radius.  We will include
five resonances $n=0,\pm 1, \pm 2$.

Within the framework of linear perturbative analysis for both the collisional damping rate and the ideal growth rate,
the net growth or damping rate ($\Gamma$) of the plasma waves in a relatively cold background plasma is
then given by
\begin{align}
  \Gamma = \Gamma_b - \Gamma_\nu.
\end{align}

\subsection{Representative runaway distribution functions}

As noted above, runaway electron distribution in
$(p,\xi)$ space can obtain an eigenmode distribution $\hat{f}_{RE}(p,\xi)$
in the avalanche phase for a fixed background plasma and parallel
electric field.  In fact, with enough collisional radial transport,
avalanche runaway electrons can even establish an eigenmode in both
momentum space $(p,\xi)$ and configuration
space.~\cite{McDevitt-etal-ppcf-2019} Here we stay with a plasma on
the magnetic axis so the avalanche runaway electron distribution
reaches a normalized eigenfunction $\hat{f}_{RE}$ in $(p,\xi).$ Let's
denote the exponentially growing runaway electron density in the
avalanche phase as $n_b(t),$ the avalanche runaway distribution
function can be simply written as
\begin{align}
f_{RE}(p,\xi,t) = n_b(t) \hat{f}_{RE}(p,\xi). \label{eq:avalanche-f}
\end{align}
This can be integrated for the runaway current density, which for
mostly relativistic electrons, is approximately $j_{RE}(t) = n_b(t) e
c,$ with $e$ the elementary charge and $c$ the light speed.

For our purpose, only one calculation of the avalanche runaway
electron eigenfunction $\hat{f}(p,\xi)$ is required for a given
background plasma, which is set by its density $n_e$, temperature
$T_e$, effective charge $Z_{eff}$, and a given parallel electric field $E.$ During the
avalanche simulated by the FPB solver, the runaway electron distribution exponentially grows over
time so that we can rescale a runaway distribution of one time frame to match
different runaway current density. We use three
types of distributions for current investigations. The first is an
example runaway distribution from the relativistic
Fokker-Planck-Boltzmann
solver~\cite{guo2017phase,guo-mcdevitt-tang-pop-2019} in the cold
disruption plasma ($T_e=10eV$) assuming $Z_{eff}=1$, $B=5.3T, n_e=10^{20}m^{-3}$ (ITER-like,
$|\omega_{ce}|/\omega_{pe}=1.65$) with an electric field in the
avalanche regime of $E= 65 E_c$. We show the
distribution below in Fig.~\ref{fig:show_distributions}(a) at
$t=3\tau_A,$ by which time the runaway current has exponentiated for
two orders of magnitude from the initial seed $0.03\times 10^{-6} $ MA/m$^2$ to $3.9\times 10^{-6} $MA/m$^2$. The second example
distribution for a larger electric field $E=200E_c$ is also calculated
at $t=1\tau_A$ with current=$5.03\times 10^{-6}$~MA/m$^2$ grown from the same runaway
seed current. The runaway distribution is
shown in Fig.~\ref{fig:show_distributions}(b).  The third case at $t=1\tau_A$ in
Fig.~\ref{fig:show_distributions}(c) builds on the second case but has
a much higher $Z_{eff}=5.$ The runaway current grows to $1.49\times 10^{-6}$~MA/m$^2$. The distributions are normalized so that the background density is unity. We also characterize three cases with the pitch integrated disbution (normalized to the same runaway current) in Fig. \ref{fig:integrated_distributions}(a) and the averaged pitch spread in Figure \ref{fig:integrated_distributions}(b). Cases 1 and 2, though with different $E/E_c$, have similar shapes of pitch averaged distributions over $p$, while case 3 with higher $Z_{eff}$ has a different shape, which has a somewhat higher averaged momentum. Case 2 with a higher $E/E_c$ has a narrower pitch spread than case 1, while case 3 with a higher $Z_{eff}$ has a broader spread than case 2. 

\begin{widetext}
\begin{figure}[H]
\centering
\includegraphics[width=0.9\textwidth]{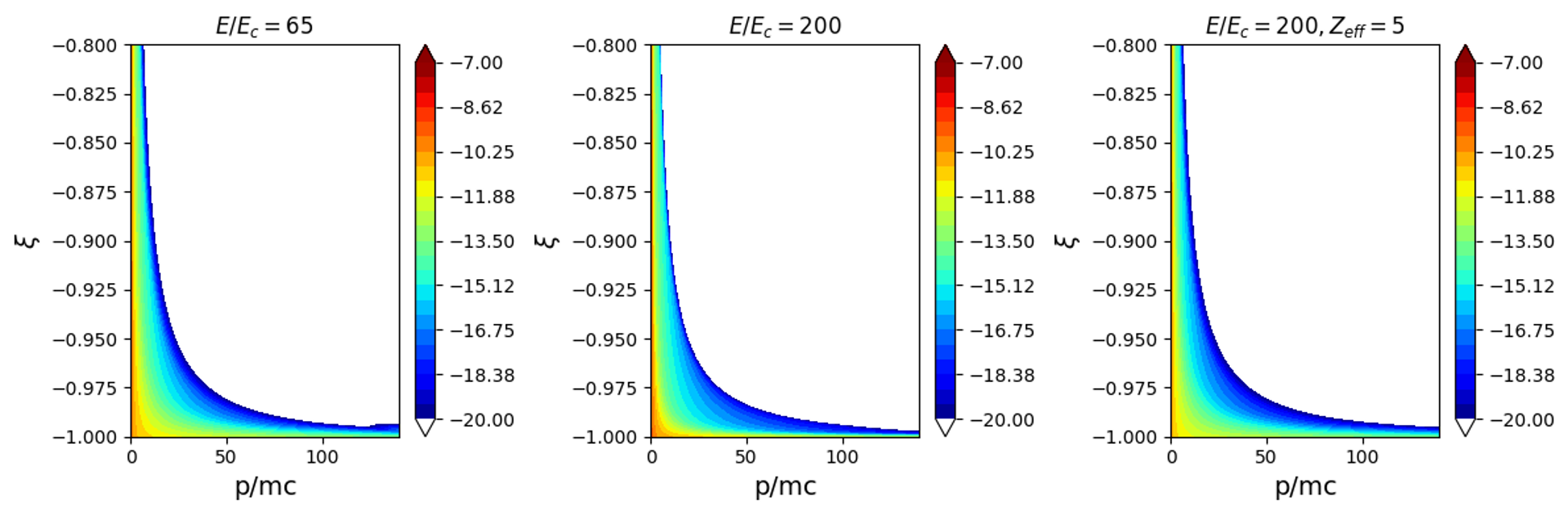}
\caption{Momentum space distribution of the three cases of calculated distributions. }
\label{fig:show_distributions}
\end{figure}
\end{widetext}

\begin{widetext}
\begin{figure}[H]
\centering
\includegraphics[width=0.9\textwidth]{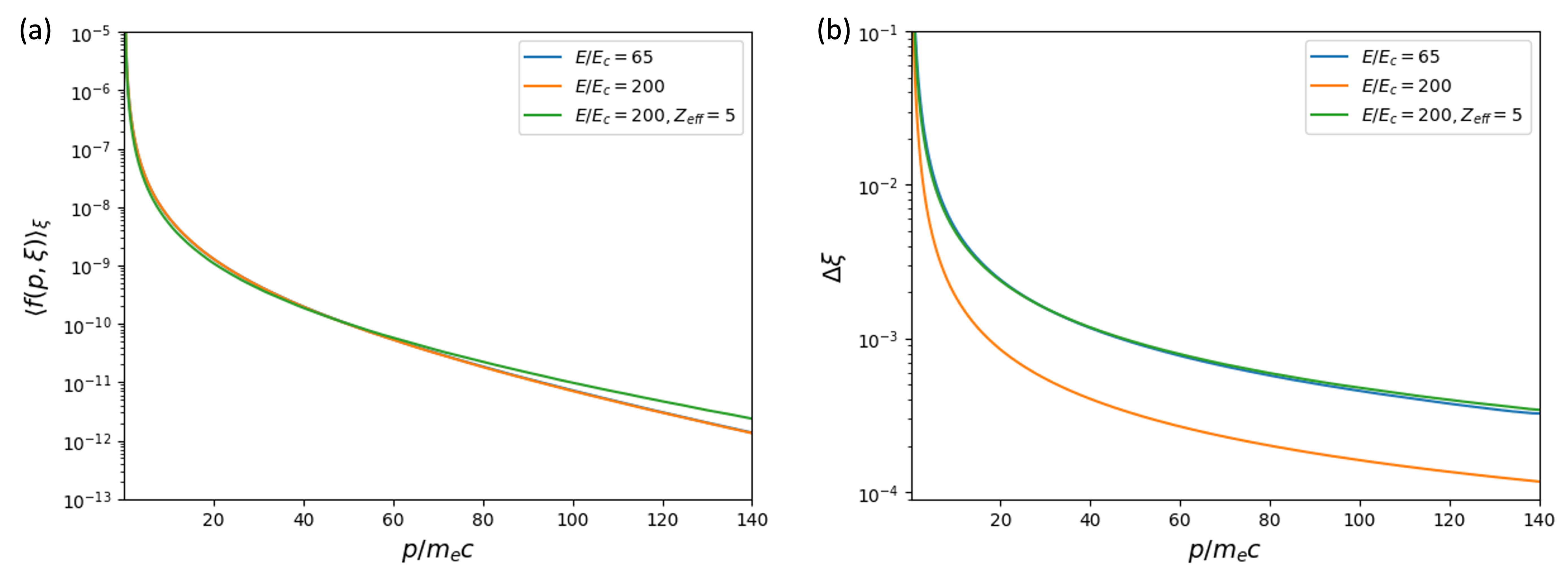}
\caption{(a) pitch-averaged distribution $\left<f(p,\xi)\right>_\xi \equiv \int_{-1}^1 f(p,\xi) d\xi$ normalized so that each case has the same runaway current 1~MA/m$^2$. Cases 1 and 2 (blue and orange lines) overlap well. The averaged momentum $\left<p\right> \equiv \frac{\int p \left<f(p,\xi)\right>_\xi 2\pi p^2 dp}{\int \left<f(p,\xi)\right>_\xi 2\pi p^2 dp}$ for the three cases are about 40,40 and 52 $m_e c$, respectively. (b)  averaged pitch spread $\left<\Delta \xi\right> \equiv \frac{\int \Delta\xi f(p,\xi) d\xi}{\int f(p,\xi)  d\xi}$, where $\Delta \xi = 1 + \xi$.}
\label{fig:integrated_distributions}
\end{figure}
\end{widetext}

\subsection{Method to scan the runaway current density dependence of runaway-driven whistler and slow-X instabilities under collisional wave damping}

In the linear perturbative analysis of the runaway-driven wave
instabilities, the ideal growth rate is linearly proportional to
$n_b(t),$
\begin{align}
\Gamma_b(f_{RE}) = n_b \hat{\Gamma}_b\left(\hat{f}_{RE}\right).  \label{eq:ideal-gamma-avalanche}
\end{align}
This is easily seen from the ideal growth rate,
Eq.~~(\ref{eq:ideal-growth-rate}), after substituting in the avalanche
distribution, Eq.~(\ref{eq:avalanche-f}), which is separable in time
and momentum space variables.  As a result, for a given background
plasma and a fixed parallel electric field, our calculation for the ideal
growth rate is done only once, for $\hat{\Gamma}_b(\hat{f}_{RE})$ from the
avalanche eigenfunction $\hat{f}_{RE}(p,\xi),$ and then the ideal growth
rate for varying runaway current density is simply scaled from
$\hat{\Gamma}_b(\hat{f}_{RE})$ via Eq.~(\ref{eq:ideal-gamma-avalanche}).
To relate the $n_b$ to $j_{RE},$ one can use the approximate relation
$j_{RE} = n_b e c,$ or for accuracy, with the light speed $c$ replaced
by the actual mean runaway parallel speed integrated from
$\hat{f}_{RE}(p,\xi).$

One can similarly scale the collisional damping rate with
respect to the electron density $n_e$, temperature $T_e$, and effective charge $Z_{eff}.$  For a reference electron density
$n_0$, temperature $T_0$ and effective charge $Z_0,$ the electron-ion collision rate $\nu_e$ (from
Eq.~(\ref{eq:nu-e})) can be written as
\begin{align}
\nu_e(n_e, T_e, Z_{eff}) = \frac{n_eZ_{eff}T_e^{-3/2}}{n_0Z_0T_0^{-3/2}} \nu_e\left(n_0, T_0, Z_0\right),
\end{align}
with
\begin{align}
  \nu_e\left(n_0, T_0, Z_0\right)
  =
  \frac{4\sqrt{2\pi} \ln\Lambda e^4}{3m_e^{1/2} T_0^{3/2}} Z_0 n_0.
\end{align}
In the linear perturbative analysis of the collisional damping rate,
$\nu_e$ enters as a linear factor in Eq.~(\ref{eq:gamma-nu}) via the
anti-Hermitian $\epsilon_{\alpha\beta}^A,$ so we have the
electron density, temperature, and effective charge scaling for the collisional damping rate,
\begin{align}
\Gamma_\nu(n_e, T_e, Z_{eff}) = \frac{n_eZ_{eff}T_e^{-3/2}}{n_0Z_0T_0^{-3/2}} \Gamma_\nu(n_0, T_0, Z_0). \label{eq:Gamma-Zeff}
\end{align}

For a background plasma of $(n_e, T_e, Z_{eff})$ and a specific
parallel electric field $E/E_c,$ we will compute the
corresponding avalanche eigenfunction $\hat{f}_{RE}(p,\xi)$ and evaluate
the ideal growth rate $\hat{\Gamma}_b(\hat{f}_{RE}).$ Note that under cold plasma assumption with a given $E/E_c$, $n_e$ and $T_e$ do not change the normalized $\hat{f}_{RE}(p,\xi)$ produced from the FPB solver. For a given runaway
current density $j_{RE},$ we next evaluate the ideal growth rate
$\Gamma_b$ from Eq.~(\ref{eq:ideal-gamma-avalanche}).
Taking the difference between $\Gamma_b$ and $\Gamma_\nu(n_e,T_e,Z_{eff}),$
we obtain the growth rate of the mode
\begin{align}
\Gamma(\omega,\mathbf{k}; j_{RE}, n_e, T_e, Z_{eff}) = \Gamma_b - \Gamma_\nu.\label{eq:growth-rate}
\end{align}
Finally we scan both $k_\parallel$ and $k_\perp,$ or equivalently
($k,\chi$), along the cold plasma wave dispersion in the whistler
and slow-X branches for the most unstable mode during the runaway
current ramp-up.

\section{Most unstable runaway-driven whistler and slow-X wave instabilities during runaway current ramp-up\label{sec:mode-scan}}

\subsection{Threshold values of runaway current density for the onset of whistlers and slow-X modes: $E/E_c$
  and $T_e$ dependence}

\begin{widetext}
\begin{figure}[H]
\centering
\includegraphics[width=0.9\textwidth]{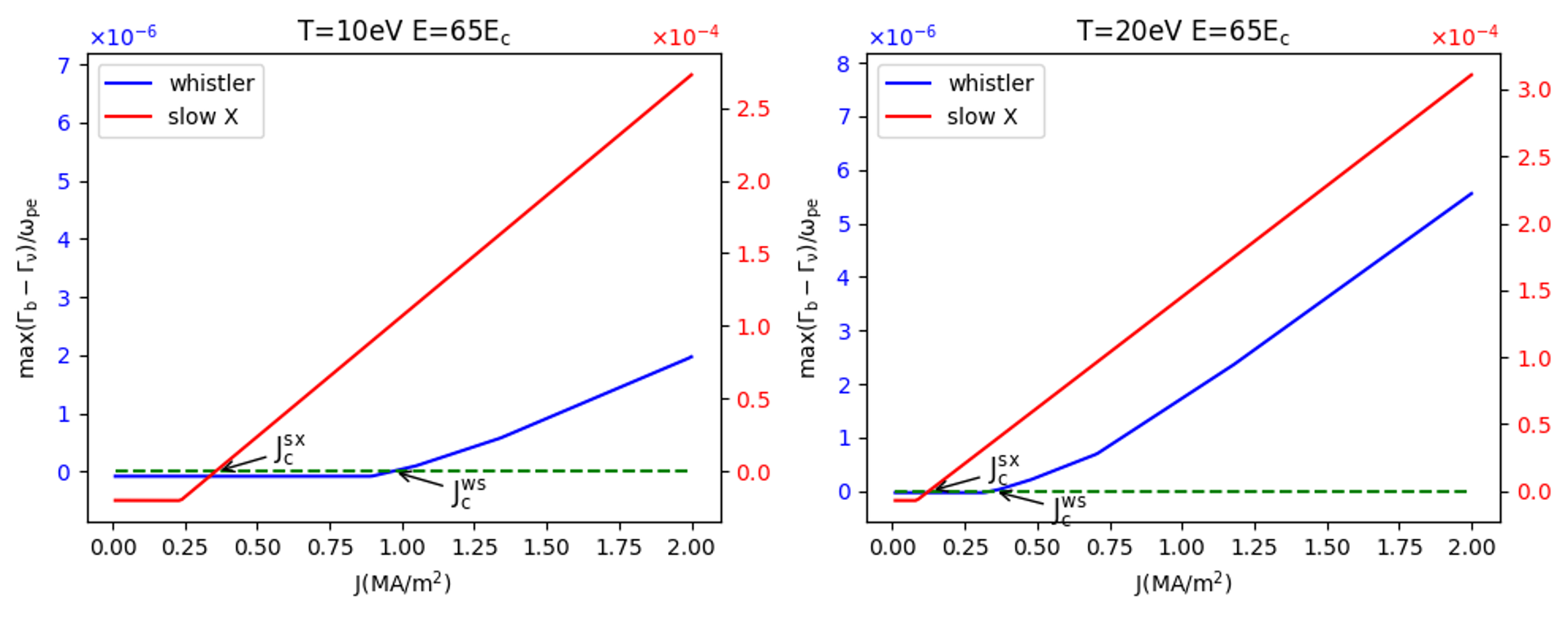}
\caption{Maximum net growth rates as a function of runaway current,
  showing that the slow-X waves get unstable before the whistlers,
  i.e. $j_c^{SX} < j_c^{WS}$. The temperature is 10eV (a) and 20eV
  (b). The scale of the whistler axis on the left is 40 times smaller
  than the slow-X axis on the right. The green dashed line denotes
  $\Gamma=0$}
\label{fig:slowx_whistler_netgrowth}
\end{figure}
\end{widetext}

As an example of our calculations, we compute the net growth rates and
plot their maximum as a function of runaway current density for the
first case in Fig.~\ref{fig:slowx_whistler_netgrowth}(a).  Both modes
are stable at low runaway current density until a critical threshold
value is surpassed.  For the whistler mode, the threshold is labeled
as $j_c^{WS},$ and for the slow-X mode, $j_c^{SX}.$ It is seen from
Fig.~\ref{fig:slowx_whistler_netgrowth}(a) that the slow-X modes can
get unstable before the whistlers, i.e. $j_c^{SX} < j_c^{WS}$, due to
the higher ideal drive rates overcoming the higher collisional
damping. This result contradicts the conventional wisdom that the
whistler waves grow more easily than the slow-X waves due to lower
collisional damping.~\cite{Aleynikov2015NucFu} In a
post-thermal-quench plasma of $T_e=10$~eV and $n_e = 10^{20}$m$^{-3}$
where we did not expect whistler instability to play a significant
role, the slow-X waves can already be strongly excited to mediate the
runaway distribution evolution. If the temperature gets higher, the
damping rate will get lower, and the threshold will shift to lower
runaway current density, e.g.,
Fig.~\ref{fig:slowx_whistler_netgrowth}(b) for $T_e=20$~eV. This is
true for even higher temperatures, which will make the damping even
weaker.  It can also be seen that at a runaway current density where
both modes are unstable, the slow-X modes have far higher growth rates
than those of the whistlers. Note that the scale of the whistler axis
on the left is 40 times smaller than the slow-X axis on the right.

\begin{widetext}
\begin{figure}[H]
\centering
\includegraphics[width=0.9\textwidth]{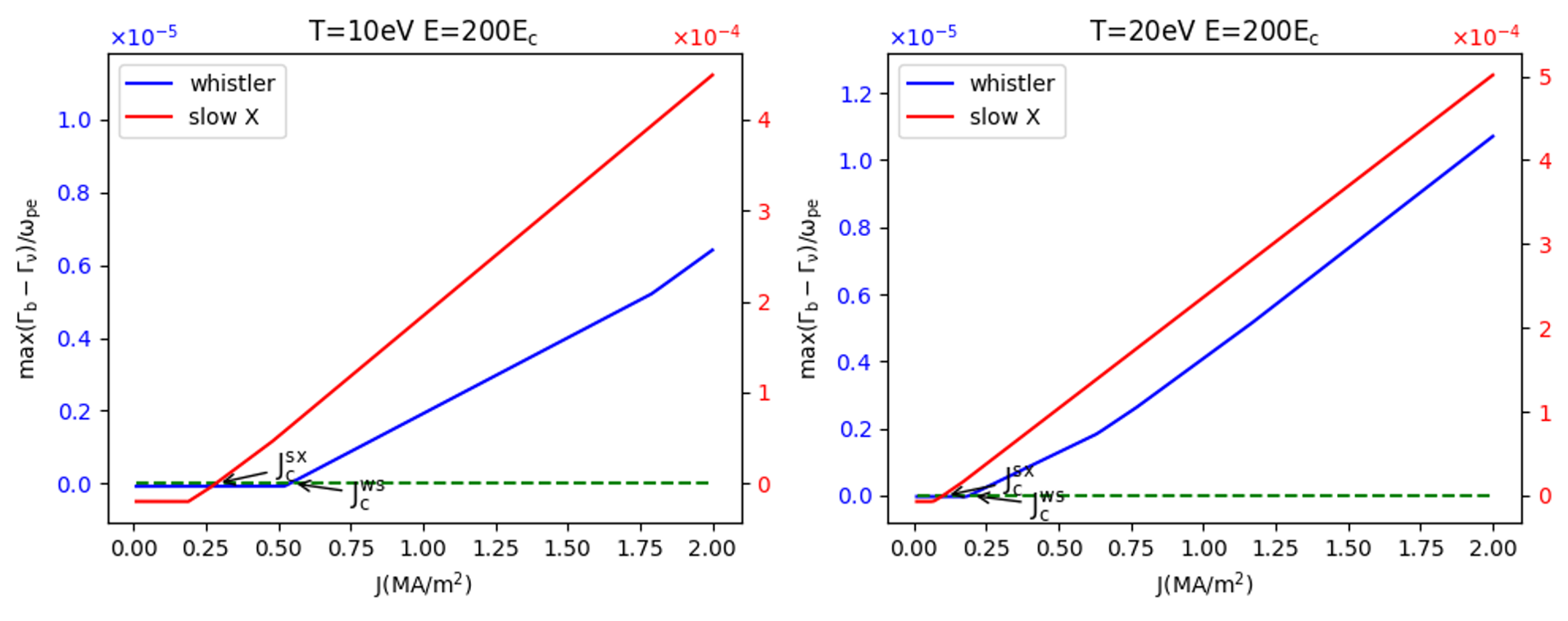}
\caption{Under $E=200E_c$, maximum net growth rates as a function of
  runaway current,
  showing $j_c^{SX} < j_c^{WS}$. The
  temperature is 10eV (a) and 20eV (b).  }
\label{fig:slowx_whistler_netgrowth_E200}
\end{figure}
\end{widetext}

A large electric field up to hundreds of $E_c$ can be realized during
disruptions in which the post-thermal-quench plasma becomes very cold.
This can lead to a more collimated runaway distribution, which
provides a stronger ideal kinetic drive.  The results for the second
case $E=200E_c,$ shown in
Fig.~\ref{fig:slowx_whistler_netgrowth_E200}, reveal similar behavior
as the $E=65 E_c$ case, in that the slow-X mode has a lower threshold
runaway current density than the whistler mode, $j_c^{SX} < j_c^{WS}.$
For the stronger ideal kinetic drive, $j_c^{SX}$ and $j_c^{WS}$ are
both down-shifted to lower values, and the gap between them also
shrinks compared with the $E=65E_c$ case.

It can be noted that we also tested on the exponential model
distribution with constant pitch spread (appendix
Eq.~\ref{eq:model-runaway-distribution}) used in
Ref.~\onlinecite{Aleynikov2015NucFu} and still get similar conclusion
of $j_c^{SX} < j_c^{WS}$.

\subsection{The most unstable whistlers and slow-X modes ($\omega, k, \chi$) versus runaway current density}

\begin{widetext}
\begin{figure}[H]
\centering
\includegraphics[width=0.9\textwidth]{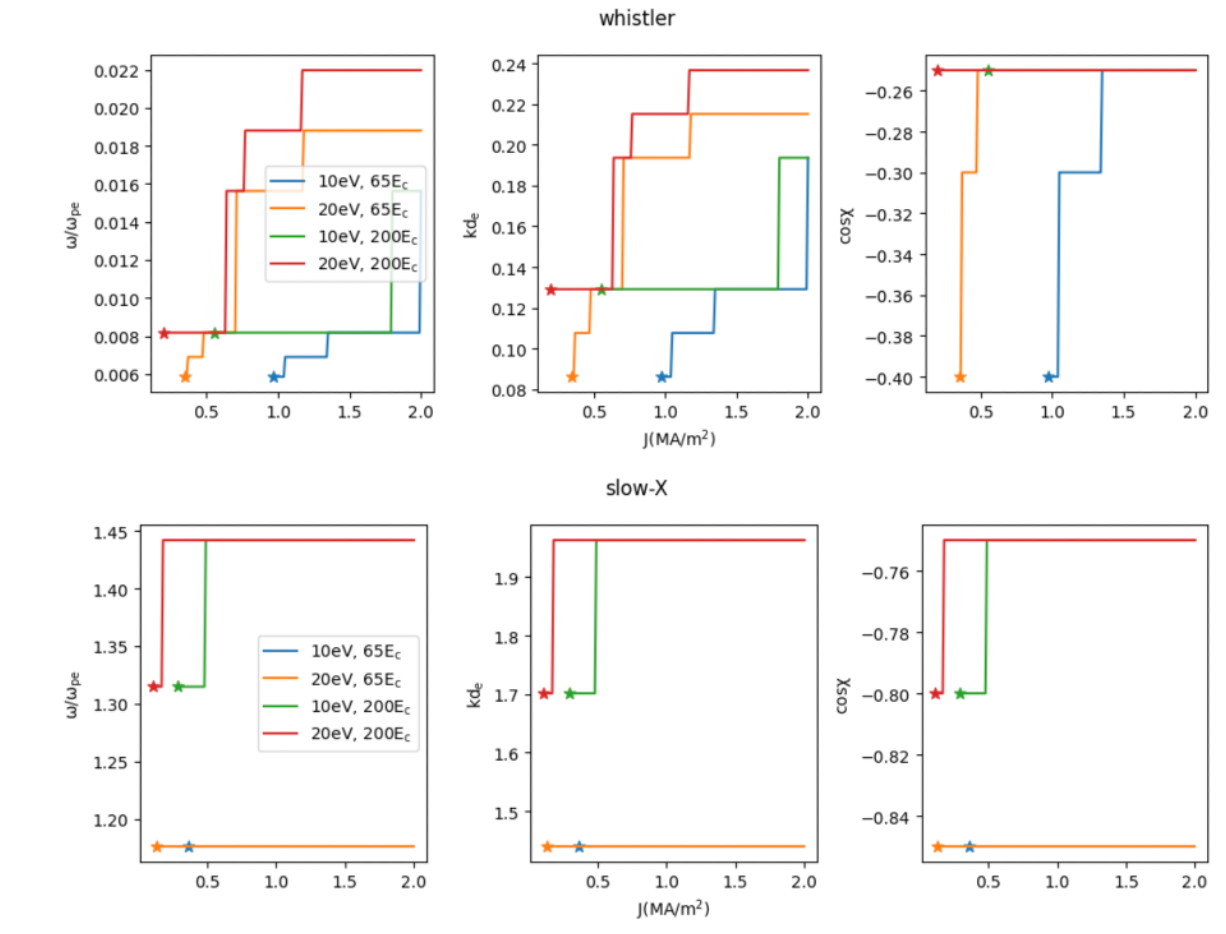}
\caption{The fastest growing modes under different parameters as a
  function of runaway current density for whistler and slow-X branches. Star
  symbols represent the threshold current density for each case. }
\label{fig:fastest_growing_modes}
\end{figure}
\end{widetext}

\begin{widetext}
\begin{figure}[H]
\centering
\includegraphics[width=0.9\textwidth]{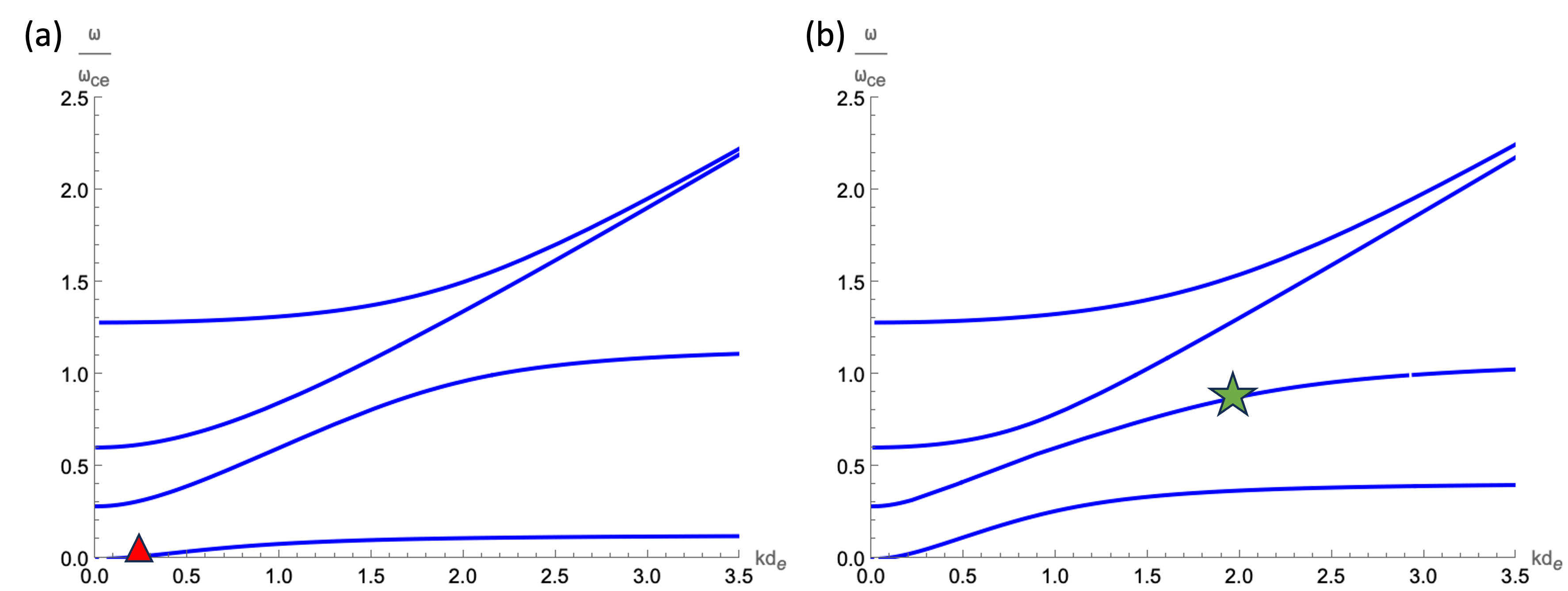}
\caption{The dispersion relation at different propagation angles are plotted to
  locate the most unstable modes for the example case of $E/E_c=200, n_e=10^{20}~\textrm{m}^{-3},
  T=20eV$ and $j_{RE}=2$~MA/m$^2$. The most unstable
  whistler is round $\chi=75^\circ$ with $kd_e=0.24$ (the red triangle in a)  resonating with parallel runaways of $\gamma\sim44$ with anomalous Doppler resonance; the most unstable slow-X mode is
  around $\chi=41^\circ$ with $kd_e=1.96$ (the green star in b), resonating with $\gamma\sim54$.  }
\label{fig:dispersion_waves}
\end{figure}
\end{widetext}

In Fig. \ref{fig:fastest_growing_modes}, we show the unstable modes $(\omega, k, \chi)$ with
the fastest net growth rates as a function of the runaway current density for the first case with $E/E_c=65$.
The most unstable slow-X mode is seen to be almost independent (with only a small relative change) of the runaway current density,
while the most unstable whistlers vary much with the runaway current density. As shown before,
a higher background electron temperature and/or a higher parallel electric field can reduce the threshold
runaway current densities (color stars). The most
unstable waves are driven at different angles $\chi$ on the dispersion
relations for slow-X and whistlers, recalling that $\cos\chi=k_\parallel/k.$
For example, at $E/E_c=200$,
$T = 20$~eV and $j_{RE} = 2$~MA/m$^2$, the most unstable slow X mode is around
$\chi=41^\circ$ with $kd_e=1.96$, while the most unstable whistler is
around $\chi=75^\circ$ with $kd_e=0.24$.
The cold plasma dispersion with these sample most unstable modes is plotted
in Fig.~\ref{fig:dispersion_waves}. The other cases look
similar, but at somewhat different $\chi$ and $k$.

\subsection{Threshold runaway current density ($j_c^{WS}, j_c^{SX}$) dependence on background electron density}

\begin{widetext}
\begin{figure}[H]
\centering
\includegraphics[width=0.9\textwidth]{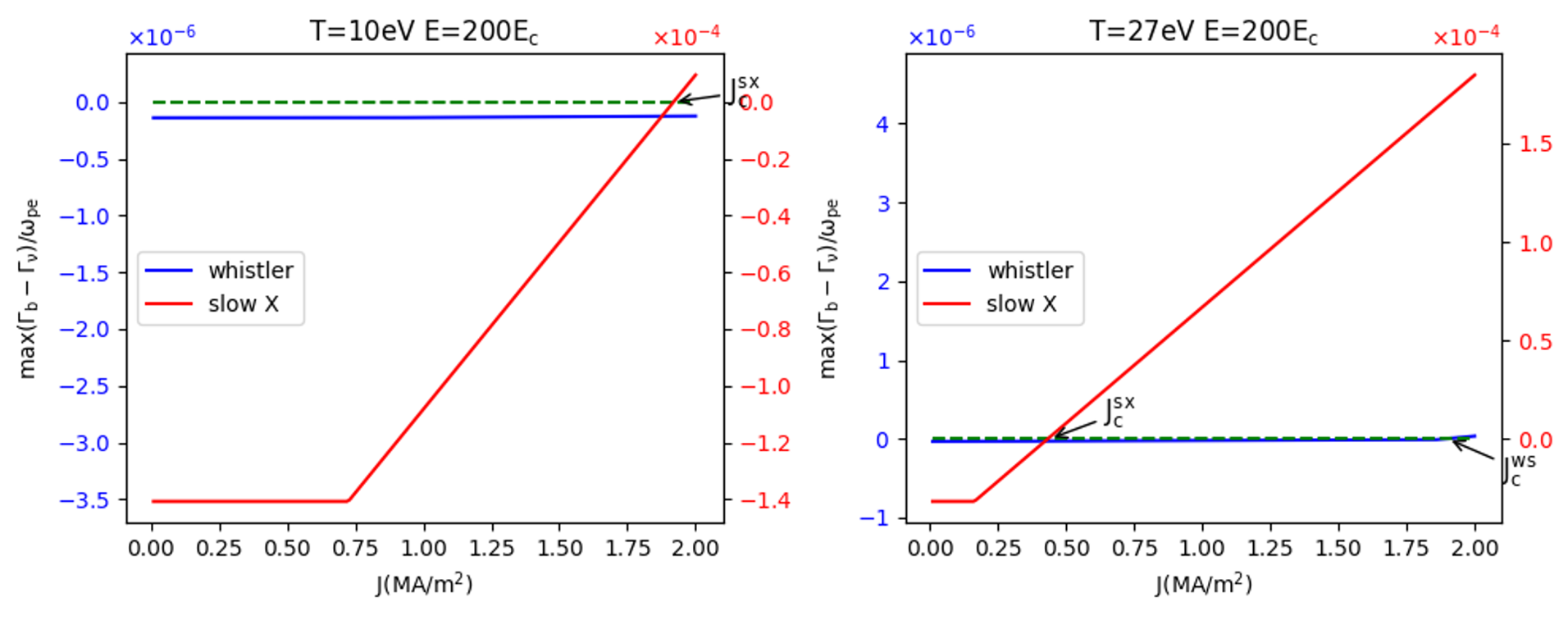}
\caption{Under $E=200E_c$ and 10 times higher density at $n_e=10^{21}$m$^{-3}$,
  maximum net growth rates as a function of runaway current. The temperature
  is 10~eV (left) and 27~eV (right). }
\label{fig:slowx_whistler_netgrowth_E200_densitytimes10}
\end{figure}
\end{widetext}

\begin{widetext}
\begin{figure}[H]
\centering
\includegraphics[width=0.9\textwidth]{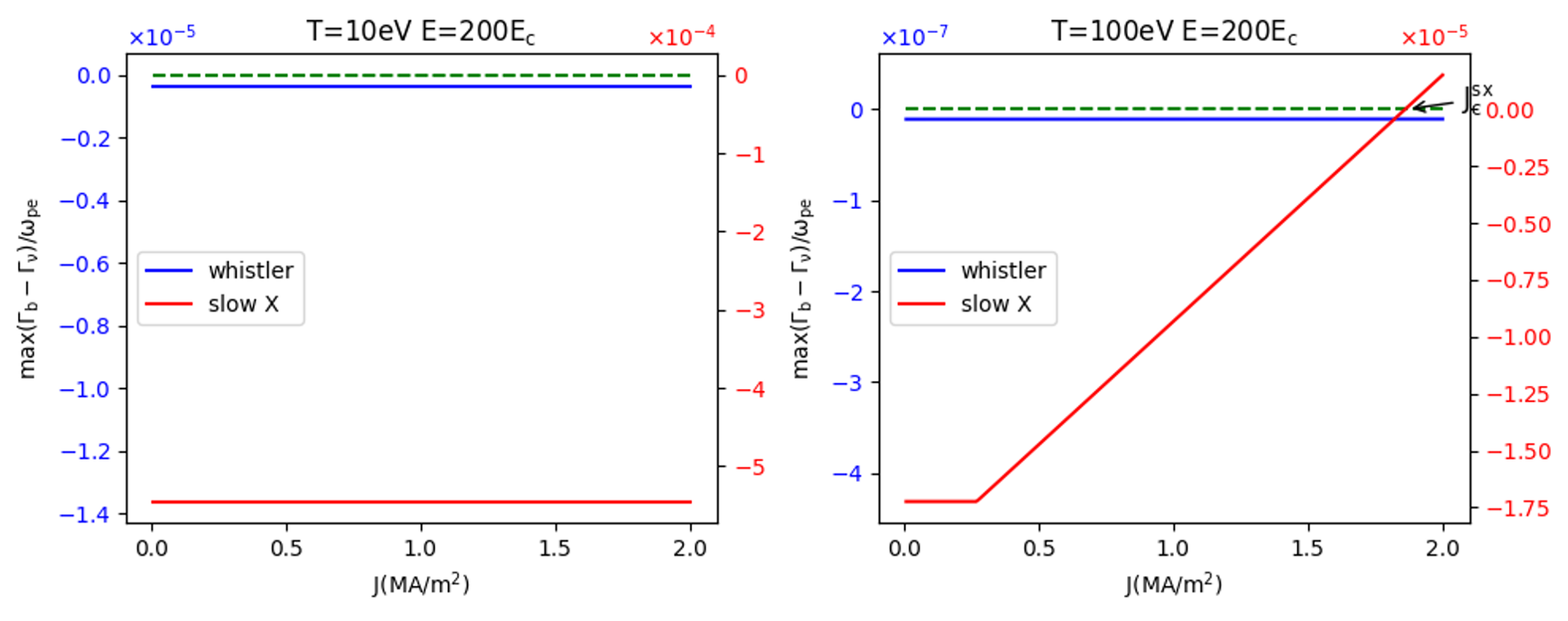}
\caption{Under $E=200E_c$ and 100 times higher density at
  $n_e=10^{22}$m$^{-3}$, maximum net growth rates as a function of
  runaway current,
  showing that no wave becomes unstable until $T_e=100$~eV for runaway
  current density up to $2$~MA/m$^2.$ The temperature is 10eV (left)
  and 100eV (right). }
\label{fig:slowx_whistler_netgrowth_E200_densitytimes100}
\end{figure}
\end{widetext}

We also investigate higher plasma density cases since the collisional damping
is proportional to the background electron density, and the density can be much higher in
mitigated disruption scenarios. Fig.~\ref{fig:slowx_whistler_netgrowth_E200_densitytimes10} shows the case of 10 times
higher density at $n_e=10^{21}$~m$^{-3}$ and
Fig.~\ref{fig:slowx_whistler_netgrowth_E200_densitytimes100} has 100 times higher density at
$n_e=10^{22}$m$^{-3}$. The threshold current density and the analogous threshold temperature
for slow-X and whistlers increase due to the increased collisional
damping. When $n_e=10^{21}$~m$^{-3}$, at 10eV, only slow-X modes are unstable
near the highest runaway current density, and the whistlers are not excited until $T_e$ is above
27~eV. When $n_e=10^{22}$~m$^{-3}$, no wave is unstable until $T_e=100$~eV. High
density has strong suppressional effects on both branches of waves.

\subsection{Threshold runaway current density ($j_c^{WS}, j_c^{SX}$) dependence on $Z_{eff}$}

\begin{widetext}
\begin{figure}[H]
\centering
\includegraphics[width=0.9\textwidth]{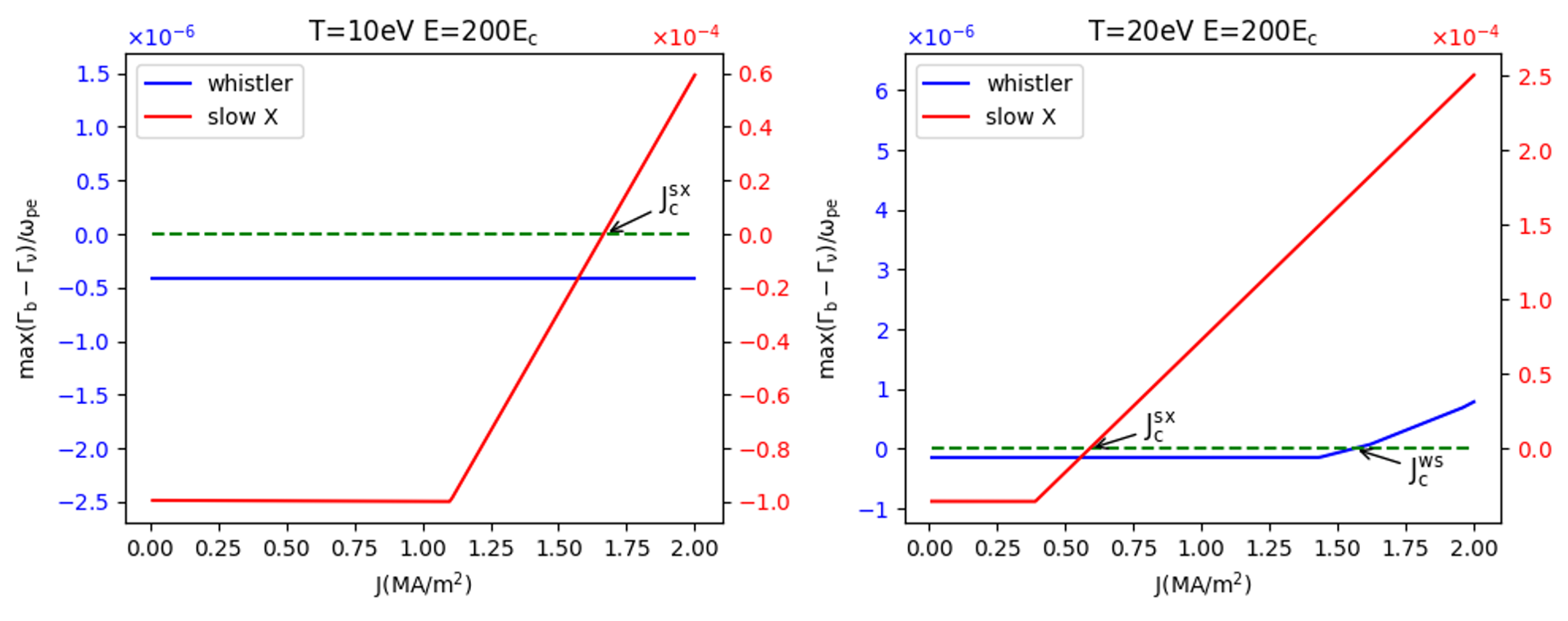}
\caption{Under $E=200E_c$ and $Z_{eff}=5$ with density
  $n_e=10^{20}$~m$^{-3}$, maximum net growth rates as a function of
  runaway current, show
  that the threshold current densities increase for both wave branches
  due to increased collisional damping and reduced ideal instability
  drive from broadened pitch distribution. The temperature is 10~eV
  (left) and 20~eV (right). See
  Fig.~\ref{fig:slowx_whistler_netgrowth_E200} for a direct comparison
  with the $Z_{eff}=1$ case.}
\label{fig:slowx_whistler_netgrowth_E200_Zeff5}
\end{figure}
\end{widetext}

Even for a fixed background electron density and $E,$ a
large $Z_{eff}$ can enhance pitch angle diffusion.
This can lead to a broader
pitch spread, which is correlated with a weaker ideal kinetic
drive for the wave instabilities.  From Eq.~(\ref{eq:Gamma-Zeff}), the
higher $Z_{eff}$ will also increase the collisional damping of the
waves linearly. We examine the case with $E/E_c=200$, $Z_{eff}=5$ in
Fig.~\ref{fig:slowx_whistler_netgrowth_E200_Zeff5}. These results can
be constrasted with the previous $Z_{eff}=1$ case of $E/E_c=200,
n_e=10^{20}$ shown in Fig.~\ref{fig:slowx_whistler_netgrowth_E200}.
At $T_e=10$~eV, the whistlers are no longer unstable for the entire
range of $j_{RE}\in (0, 2]$~MA/m$^2,$ and the critical runaway current
  density for exciting the slow-X mode is pushed up to
  $j_C^{SX}\approx 1.7$~MA/m$^2.$ At $T_e=20$~eV, the whistlers are
  excited at a higher runaway current density ($j_C^{WS} \approx
  1.6$~MA/m$^2$ in Fig.~\ref{fig:slowx_whistler_netgrowth_E200_Zeff5})
  as opposed to the $Z_{eff}=1$ case of $j_C^{WS} \approx 0.22$~MA/m$^2$ in
  Fig.~\ref{fig:slowx_whistler_netgrowth_E200}.  The slow-X is also
  excited at a higher $j_C^{SX} \approx 0.6$~MA/m$^2.$ The gap
  between $j_C^{SX}$ and $j_C^{WS}$ becomes much wider in the
  $Z_{eff}=5$ case.
  
\section{Conclusions}\label{sec:conclusion}

The runaway avalanche distribution is separable in time and momentum
space, and thus provides a most convenient case study to examine the
qualitative and quantitative influence of plasma density,
temperature and effective charge on the critical runaway electron densities for exciting
the two branches of whistler and slow-X modes in a post-thermal-quench
collisional plasma.  What we find is that the slow-X modes -- which tend
to have a higher ideal drive than whistlers but suffer from stronger
collisional damping for their higher wave numbers and frequencies -- are
more easily excited by a runaway electron distribution out of an
avalanche. This qualitative trend holds for modestly large and fairly
large $E/E_c.$ Quantitatively, the gap between the critical
runaway electron density to excite whistlers ($j_C^{WS}$) and slow-X
($j_C^{SX}$) becomes smaller for larger $E/E_c,$ which tends
to produce a more collimated runaway beam.  The most
unstable slow-X modes for different runaway current densities appear
to have the similar wave characteristics ($\omega, k, \chi),$ but the
most unstable whistlers can vary significantly in wave characteristics
for different runaway current densities. Higher background electron
density increases the collisional damping linearly.  For
$n_e=10^{21}$~m$^{-3},$ which is ten times the normal ITER density at
steady-state operation, whistlers are stable at $T_e=10$eV but can
become unstable at $T_e=27$~eV at very higher runaway current density
(about 1.9~MA/m$^{-2}$).  The slow-X modes require a very high
runaway current density (about 1.9~MA/m$^{-2}$) to be excited at
$T_e=10$~eV, but becomes a rubustly unstable with $j_C^{SX}\approx
0.5$~MA/m$^{-2}$ when $T_e$ is raised to 20~eV.  At 100 times the
normal ITER density, which has $n_e=10^{22}$~m$^{-3},$both whistlers
and slow-X are strongly collisionally damped, and it requires $T_e$ to
reach 100~eV for the slow-X mode to become unstable at the high
runaway current density of $\sim 1.9$~MA/m$^{-2}.$

When high-Z impurities are injected for disruption mitigation, the
partial screening effect would give rise to enhanced pitch angle
scattering, resulting in an avalanche distribution that has a broader
pitch spread.  This is modeled by a higher $Z_{eff},$ where we observe
much weaker ideal instability drive and stronger collisional
damping. The net effect is an even greater separation in the stability of
the whistler and slow-X modes. For example, at $T_e=20$~eV,
$E/E_c=200,$ and $n_e=10^{20}$~m$^{-3},$ both $j_C^{SX}$ and
$j_C^{WS}$ are upshifted with $Z_{eff}=5$, but the gap between them becomes much wider
compared with a pure hydrogen plasma at $Z_{eff}=1.$

To conclude, one can expect that slow-X modes are much easier to
excite by runaway avalanche distributions in a post-thermal-quench
plasma.  Even if both slow-X and whistlers are excited, for example,
by a high enough runaway current density, or a low enough plasma
density, or a high enough plasma temperature, the most unstable slow-X
mode will have far higher growth rate than the most unstable
whistlers. For self-mediation of runaway electrons by self-excited
plasma wave instabilities, there is a compelling case to recognize the slow-X modes as an important area of focus.

\begin{acknowledgments}
Qile Zhang acknowledges helpful conversations with Qi Tang.
We thank the U.S. Department of Energy Office
of Fusion Energy Sciences and Office of Advanced Scientific Computing
Research for support under the Tokamak Disruption Simulation (TDS) and
SCREAM Scientific Discovery through Advanced Computing (SciDAC)
projects, the Base Fusion Theory Program, and the General
Plasma Science program, at Los Alamos National Laboratory (LANL)
under contract No. 89233218CNA000001.  This research used resources of
the National Energy Research Scientific Computing Center, a DOE Office
of Science User Facility supported by the Office of Science of the
U.S. Department of Energy under Contract No. DE-AC02-05CH11231 using
NERSC award FES-ERCAP0028155 and the Los Alamos National Laboratory
Institutional Computing Program, which is supported by the
U.S. Department of Energy National Nuclear Security Administration
under Contract No. 89233218CNA000001.
\end{acknowledgments}

%

\appendix
\renewcommand{\thefigure}{A.\arabic{figure}}
\setcounter{figure}{0}

\section{Benchmark of the collisional damp rates and ideal growth rates with Ref.~\onlinecite{Aleynikov2015NucFu} and  Ref.~\onlinecite{Komar2012JPhCS} \label{sec:benchmark}} 

\begin{figure}[H]
\centering
\includegraphics[width=0.4\textwidth]{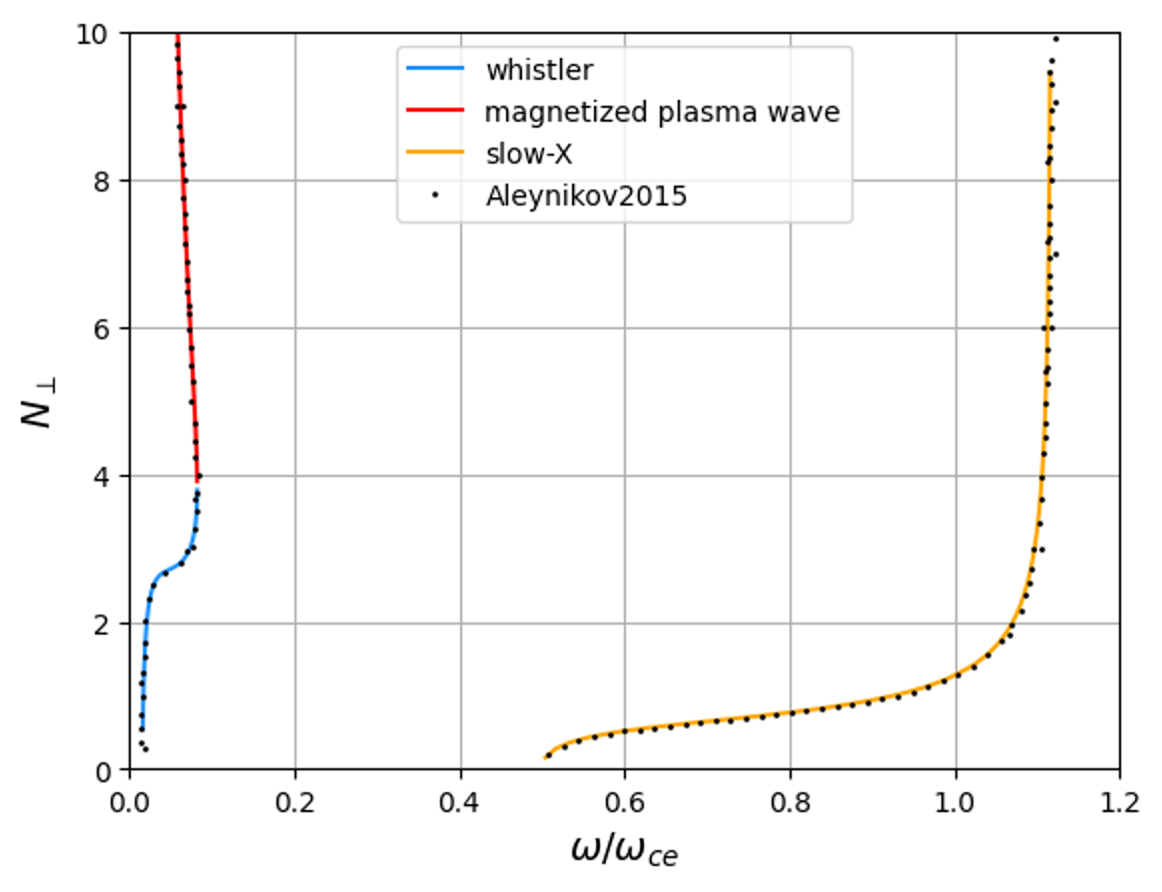}
\caption{ Benchmark case (I): Transverse refractive index $N_\perp$ and
  frequency window for waves driven via the anomalous Doppler
  resonance by electrons with $\gamma=20.$ The ratio of
  $\omega_{pe}/|\omega_{ce}|=0.5$ The solid lines are from the current
  calculation and the dots are data from Fig.~1 in
  Ref.~\onlinecite{Aleynikov2015NucFu}.}
\label{fig:benchmark_fig1}
\end{figure}

\begin{figure}[H]
\centering
\includegraphics[width=0.4\textwidth]{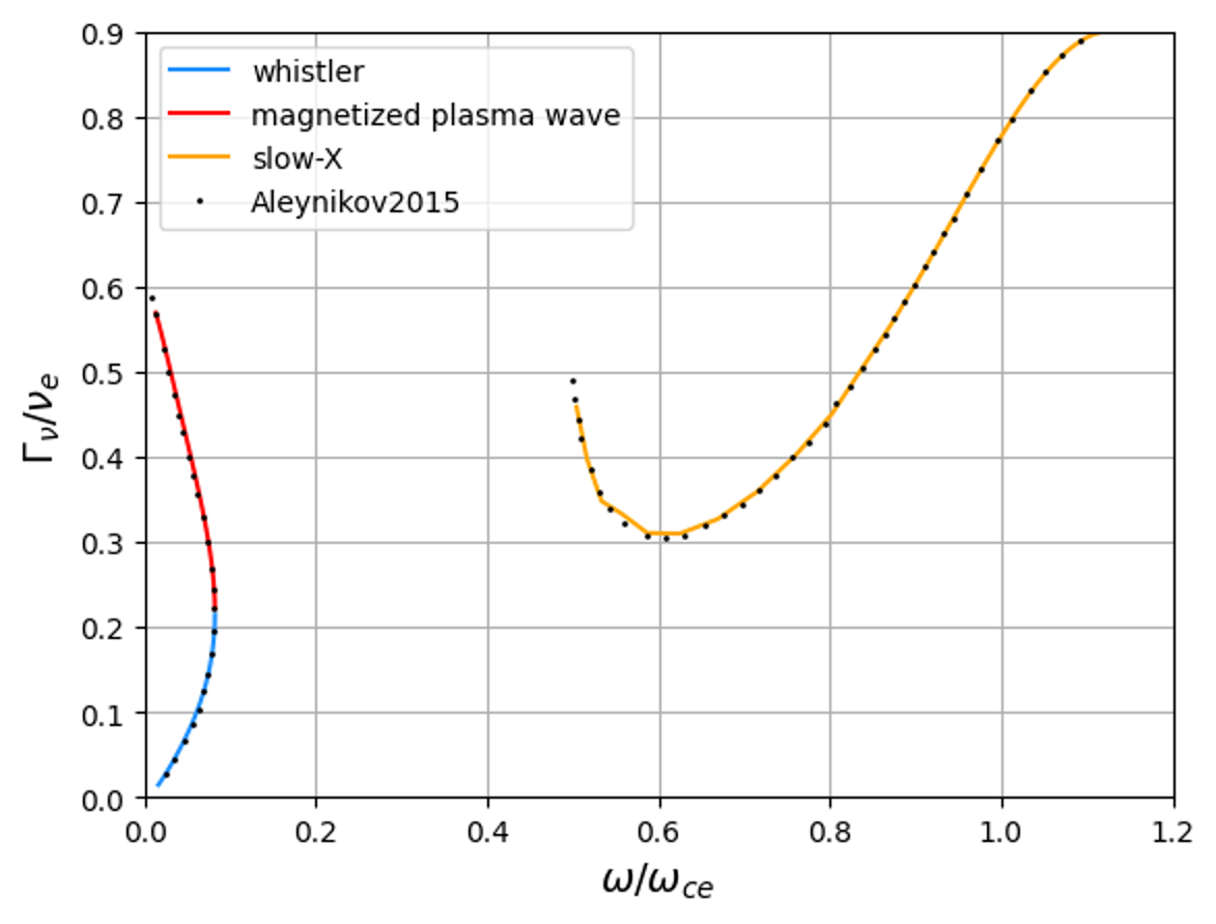}
\caption{Benchmark case (I): Collisional damping rates for whistler-magnetized-plasma
  and slow-X waves driven via anomalous Doppler resonance by electrons
  with $\gamma=20.$ The ratio of $\omega_{pe}/|\omega_{ce}|$ is 0.5.
  The color lines are from the current calculation, and the dots are data
  from Fig.~2 of Ref.~\onlinecite{Aleynikov2015NucFu}. }
\label{fig:benchmark_damping}
\end{figure}

We were concerned by the accuracy of the linear dispersion calculation
of the runaway-driven whistler and slow-X modes in a collisional
plasma, and have gone through some benchmark studies with the results
reported in Ref.~\onlinecite{Aleynikov2015NucFu}.  Two specific cases are
considered. The first concerns with the collisional damping rate of
the whistler and slow-X waves that are in resonance with a runaway
electron of Lorentz factor $\gamma=20$ and perfectly aligned with the
magnetic field, namely unity pitch $\xi=1.$ The resonance condition of
the primary anomalous Doppler resonance ($n=1$) constrains the transverse
refractive index as a function of wave frequency for both branches of
waves, which were shown in Fig.1 of
Ref.~\onlinecite{Aleynikov2015NucFu}. The corresponding collisional damping
rate, normalized by the electron collision rate, as a function of the
wave frequency, is shown in Fig.~2 of Ref.~\onlinecite{Aleynikov2015NucFu}.
This benchmark uses the ratio  $|\omega_{ce}|/\omega_{pe}=2$ with infinite ion-electron mass ratio.
Our code produces identical results for both the resonant wave
dispersion $(\mathbf{N}_\perp,\omega)$ and the collisional damping rate
$\Gamma_\nu,$ compared with those in
Ref.~\onlinecite{Aleynikov2015NucFu}. These are shown in Fig.~\ref{fig:benchmark_fig1} and
Fig.~\ref{fig:benchmark_damping}. Note that in these benchmark figures we adopt the notation from Ref.~\onlinecite{Aleynikov2015NucFu} of splitting the whistler branch into the whistler (the low
  $N_\perp$ portion in Fig.~\ref{fig:benchmark_fig1}) and magnetized plasma waves (the high
  $N_\perp$ portion in Fig.~\ref{fig:benchmark_fig1}).

\begin{figure}[H]
\centering
\includegraphics[width=0.45\textwidth]{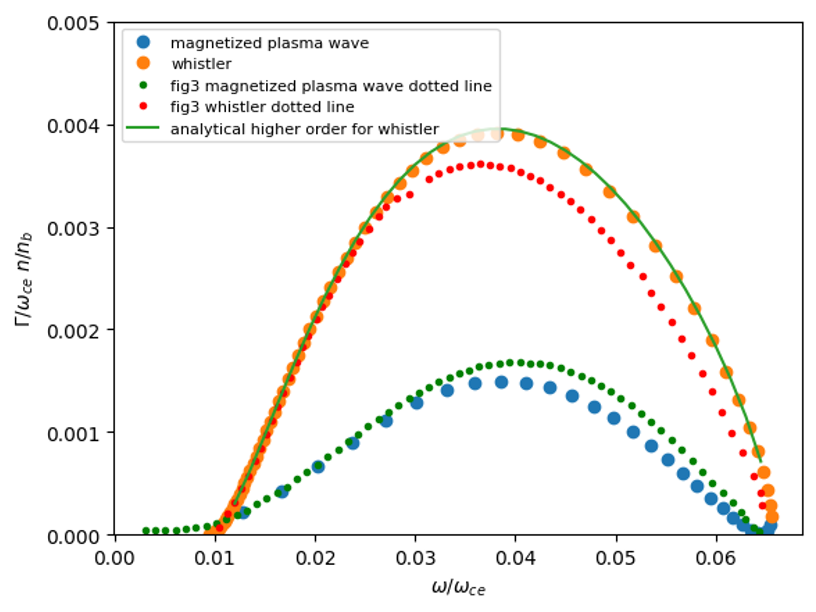}
\caption{Benchmark case (II): Ideal growth rates of the whistler and
  magnetized plasma waves driven by a runaway model distribution of
  Eq.~(\ref{eq:model-runaway-distribution}) with $p_0$ corresponding
  to $\gamma=25$ and $\theta_0=0.1.$ The ratio of
  $\omega_{pe}/|\omega_{ce}|$ is 0.5.  The data points in large solid
  circles are from current calculation, while those in small solid
  circles are from Fig.~3 of
  Ref.~\onlinecite{Aleynikov2015NucFu}. There are appreciable
  discrepancies between the two. For the whistler modes, current
  calculation is in good agreement with analytical results from a
  higher order expansion in $\theta.$}
\label{fig:benchmark_fig3}
\end{figure}

The second case concerns the ideal growth rate calculation of the whistler-to-magnetized-plasma-wave branch
using a model exponential runaway distribution of the form~\cite{Aleynikov2015NucFu}
\begin{align}
  F_b(p,\theta) = \frac{n_b}{2\pi} \frac{\exp\left(-\frac{p}{p_0}\right)}{p_0p^2}
  \frac{2\exp\left(-\frac{\theta^2}{\theta_0^2}\right)}{\theta_0^2},
 \label{eq:model-runaway-distribution}
\end{align}
to benchmark the calculation in Fig.3 of
Ref.~\onlinecite{Aleynikov2015NucFu}, where $p_0=25$,
$\theta_0=0.1$. Ref.~\onlinecite{Aleynikov2015NucFu} considers only
the primary anomalous resonance $n=1$.  Since $\int
F_b(p, \theta) \, d^3p \approx n_b$, the original equation Eq. (31) of
Ref.~\onlinecite{Aleynikov2015NucFu} appears to contain a typo,
positioning the coefficient 2 of the $\theta$ term into the
denominator.  In Fig.~\ref{fig:benchmark_fig3}, we show the current
calculation of the normalized ideal growth rates, in large solid
circles, for both the whistler and magnetized plasma wave as a
function of wave frequency $\omega.$ The whistlers have a much higher
growth rate (large orange dots) compared with the magnetized plasma
waves (large blue dots).  But the results differ from Fig.3 of
Ref.~\onlinecite{Aleynikov2015NucFu} (small red and green dots) in
that the current calculation produces somewhat higher growth rates for
the whistler modes and lower growth rates for the magnetized plasma
waves.

To gain further insights on the discrepancy, we also benchmark with a
higher-order analytical solution with respect to $\theta$ for the
whistler mode under consideration.  The analysis begins by recalling
Eq.~\ref{eq:ideal-growth-rate} in our main text (similar to Eq.~(21)
of Ref.~\onlinecite{Aleynikov2015NucFu}) for the ideal instability
drive.  Recall that Ref.~\onlinecite{Aleynikov2015NucFu} only
considers the anomalous resonance $n=1$ in this calculation.  We first
integrate over $p$ to remove the delta function and all $p$ in the
integral become a function of $\theta$ corresponding to Lorentz factor
\begin{align}
\gamma = 
\frac{
    k_\parallel^2 c^2 \cos^2\theta + \omega_c^2
}{
    \omega \omega_c + 
    \sqrt{
        \omega^2 \omega_c^2 + 
        \left(k_\parallel^2 c^2 \cos^2\theta - \omega^2\right)
        \left(k_\parallel^2 c^2 \cos^2\theta + \omega_c^2\right)
    }
}.
\end{align}
This leaves only an integration over $\theta$, which can be
numerically integrated over $\theta$ using Mathematica to verify our
numerical integration results, but we also further pursue an
analytical verification below. Next, we seek a higher order analytical
solution by keeping the extra higher-order components (by $\theta^2$
higher) in the integrand and truncating at $\theta^3$.  Integrating
\(\frac{\partial F_\text{b}}{\partial \theta}\) by part to remove the
$\frac{\partial}{\partial \theta}$, we find the results can be
simplified to a form of
\[
\int_{0}^{\pi}\bigl(C_{1}\,\theta + C_{2}\,\theta^{3}\bigr)\,e^{-C_{3}\,\theta^{2}}\;d\theta,
\]
 which is still analytically integrable, although $C_1, C_2, C_3$ are
 complicated functions of the wave modes. This higher-order analytical
 solution (represented by the green solid line) is in good agreement
 with our numerical results for the whistler modes (large orange
 dots).  It should be noted that such a higher-order expansion in
 $\theta$ may not apply to the magnetized plasma waves that correspond
 to nearly perpendicular waves, since the low-order expansion of
 Bessel function used in the solution would be invalidated because of
 the argument $k_\perp \rho \gg 1$.  It remains unclear what may have
 contributed to the discrepancy between our results, which were
 obtained separately from analytical analysis and numerical
 integration, and those computed in
 Ref.~\onlinecite{Aleynikov2015NucFu}.

\begin{figure}[H]
\centering
\includegraphics[width=0.45\textwidth]{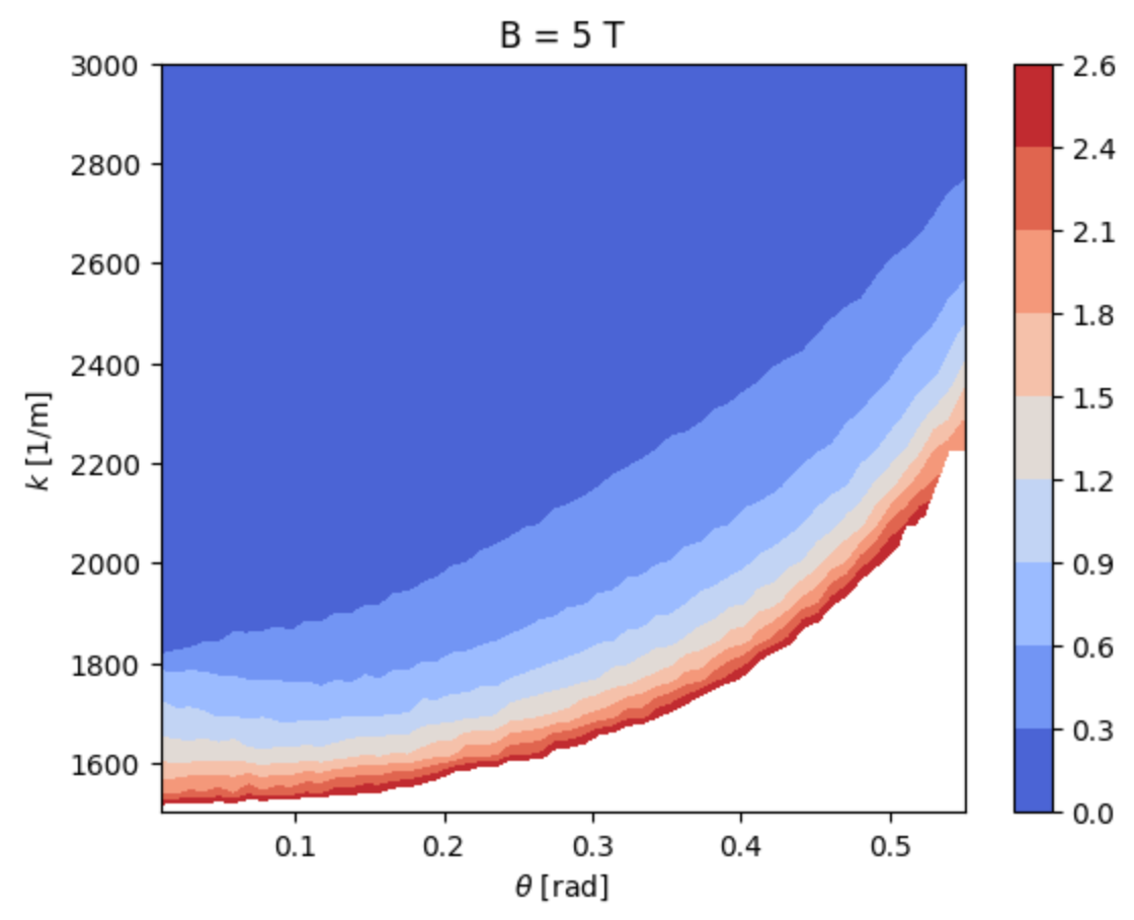}
\caption{Benchmark case (III): Contours of ideal growth rates of
  slow-X modes to benchmark a case in the Fig. 4(b) of
  Ref.~\onlinecite{Komar2012JPhCS} with parameters $B = 5$~T and a
  model distribution for near-critical runaway distribution. As
  numerical integrations, our results are qualitatively close to
  theirs, suggesting our code is producing reasonable results. Here
  $\theta$ instead of $\chi$ represents the wave propagation angles. }
\label{fig:benchmark_Komar2012_fig4}
\end{figure}

 Since Ref.~\onlinecite{Aleynikov2015NucFu} did not calculate slow-X
 modes for their ideal growth rate, we benchmark with another work on
 slow-X modes from Ref.~\onlinecite{Komar2012JPhCS}. Fig.~4 of
 Ref.~\onlinecite{Komar2012JPhCS} calculates the ideal growth rate of
 slow-X modes that accounts for both the anomalous Doppler and
 Cherenkov resonances on a near-critical ($E/E_c\gtrsim1$) model
 runaway distribution. Our results in
 Fig.~\ref{fig:benchmark_Komar2012_fig4} show qualitative agreement
 with their results in Fig.~4(b) of Ref.~\onlinecite{Komar2012JPhCS}
 as numerical integrations. Note that only in this figure we use
 $\theta$ instead of $\chi$ to represent the wave propagation angles,
 to be consistent with Ref.~\onlinecite{Komar2012JPhCS}. We discuss
 below a caveat on the integration process. As shown in
 Ref.~\onlinecite{Komar2012JPhCS,Komar2013pop}, the runaway
 distribution is given by
 \begin{widetext}
\[
f_r(p_\parallel, p_\perp) = \frac{A}{p_\parallel^{(C_s - 2)/(\alpha - 1)}} \exp\left( -\frac{(\alpha + 1)p_\perp^2}{2(1 + Z)p_\parallel} \right) 
{}_1F_1\left( 1 - \frac{C_s}{\alpha + 1}, 1; \frac{(\alpha + 1)p_\perp^2}{2(1 + Z)p_\parallel} \right),
\]
 \end{widetext}
where
\[
C_s = \alpha - \frac{(1 + Z)}{4}(\alpha - 2)\sqrt{\frac{\alpha}{\alpha - 1}},
\]
\( Z=1 \) is the effective ion charge, $\alpha=E/E_c=1.3$ and \(
   {}_1F_1 \) is the confluent hypergeometric function.  Since the
   integration over the full momentum space is divergent, the
   normalization factor A is given by a finite integration range \[
   \int_0^\infty dp_\perp\, 2\pi p_\perp
   \int_{p_c}^{p_{\parallel{max}}} dp_\parallel\, f_r(p_\parallel,
   p_\perp) = 1,
\] where the critical momentum $p_c=1.73$, $p_{\parallel{max}}=5$
 (normalized to $m_e c$)~\cite{Komar2012JPhCS,Komar2013pop}. However,
the wave modes with strong growth rates in the figure are in  resonance
with particles at $p\gg5$. To calculate the growth rates, the
distribution must be integrated to a momentum upper bound
$p_{max}\gg5$, but the upper bound being used in
Ref.~\onlinecite{Komar2012JPhCS,Komar2013pop} appears not given.  In
our growth rate calculation, we choose to integrate to $p_{max}=100$.
 
\end{document}